\documentstyle[graphicx]{mn}

\newif\ifAMStwofonts
\AMStwofontstrue

\def\mkn{{Mkn 766}}

\def\xmm{{\it XMM-Newton}}
\def\chandra{{\it Chandra}}

\def\et{{et al.\ }}

\def\rosat{{\it ROSAT}}

\def\asca{{\it ASCA}}
\def\sax{{\it BeppoSAX}}

\newcommand{\ls}{\mathrel{\hbox{\rlap{\hbox{\lower4pt\hbox{$\sim$}}}\hbox{$<$}}}}
\newcommand{\gs}{\mathrel{\hbox{\rlap{\hbox{\lower4pt\hbox{$\sim$}}}\hbox{$>$}}}}

\def\arcs{{\hbox{$^{\prime\prime}$}}}

\def\Msun{\hbox{$\rm ~M_{\odot}$}}

\def\H0{{\rm ~km~s^{-1}~Mpc^{-1}}}

\def\et{{et al.}}

\title[Fe K emission and absorption in \mkn]
        {Fe K emission and absorption features 
	in \xmm\ spectra of \mkn\ - evidence for reprocessing in  flare ejecta}
\author[K.A.Pounds \et]
        {K.A.Pounds,$^{1}$
	J.N.Reeves,$^{1,2}$
	K.L.Page,$^{1}$
	G.A.Wynn,$^{1}$
         and P.T.O'Brien,$^{1}$\\
$^1$ Department of Physics and Astronomy, University of Leicester,
Leicester, LE1 7RH, UK\\
$^2$ Laboratory for High Energy Astrophysics, NASA Goddard Space Flight Center, Greenbelt, MD 20771, USA\\}
\date{Accepted ; Submitted }
\pagerange{\pageref{firstpage}--\pageref{lastpage}}
\pubyear{2003}
\begin{document}
\maketitle
\label{firstpage}

\begin{abstract}

We report on the analysis of a long \xmm\ EPIC observation in 2001 May
of the Narrow Line Seyfert 1 galaxy \mkn. The 3--11 keV spectrum exhibits a moderately steep power law
continuum, with a broad emission line at
$\sim$6.7 keV, probably blended with a
narrow line at $\sim$6.4 keV, and a broad absorption trough above
$\sim$8.7 keV. We identify both broad spectral features with reprocessing in He-like Fe. 
An earlier \xmm\ observation of \mkn\ in 2000 May, when the source was a factor $\sim$2 
fainter, shows a similar broad emission line, but with a slightly flatter power law and
absorption at a lower energy.
In neither observation do we find a requirement for
the previously reported broad `red wing' to the line and hence of
reflection from the innermost accretion disc. 
More detailed examination of the longer \xmm\ observation reveals evidence for rapid spectral variability in the Fe K band, apparently 
linked with the 
occurrence of X-ray `flares'. A reduction in the emission line strength and increased high energy absorption during the X-ray flaring 
suggests that these transient effects are due to highly ionised ejecta associated with the flares. Simple scaling from the flare avalanche
model proposed for the luminous QSO PDS 456 (Reeves \et 2002) confirms the feasibility of coherent flaring being the cause of the strong
peaks seen in the X-ray light curve of \mkn.

\end{abstract}


\section{Introduction}

Perhaps the most striking recent development in X-ray studies of AGN has been the observation, from high resolution
grating spectra with \chandra\ and \xmm, of complex absorption indicating circumnuclear (often outflowing) matter existing in a wide
range of
ionisation states (eg Sako \et\ 2001, Kaspi \et\ 2002). To date, however, it has generally been assumed that this so-called `warm 
absorber' was
essentially transparent in the `Fe K spectral band' above $\sim$6 keV. This is important as,
following its recognition over a decade ago as a common feature in the X-ray spectra of AGN (Pounds \et\ 1990), 
the Fe K emission line has assumed a diagnostic potential comparable to that of H Ly-alpha.  
The subsequent observation of a relativistically broadened Fe K line in the Seyfert galaxy MCG-6-30-15 
(Tanaka \et\ 1995) offered a unique probe of strong gravity (Fabian \et\ 2000); moreover, the relativistic broad 
line was 
reported to be a common property of
Seyfert 1 galaxies in \asca\ spectra (Nandra \et\ 1997). 

Although,  
with a few notable exceptions (eg MCG-6-30-15;
Wilms \et\ 2002; Fabian \et\ 2002), the comparable but more sensitive \xmm\ EPIC observations have so far failed to
confirm the \asca\ results, in particular the strong `red wing' to the
Fe K line, the Fe K spectrum remains of high diagnostic potential (eg Reynolds and Nowak 2002).
The revised picture arising from \xmm\ and \chandra\ observations includes
the frequent
detection of a narrow emission line at $\sim$6.4~keV (Yaqoob \et\ 2001, Kaspi \et\ 2001, O'Brien \et\ 2001,
Gondoin \et\ 2001,Pounds \et\ 2003),
presumed to have
a distinct origin, away from the inner disc, together - in several
cases - with a broad emission feature at $\sim$ 6.4-7 keV, indicative of emission from ionised matter 
(Reeves \et\ 2001, 
Pounds \et\ 2001, Matt \et\ 2001). 

In this paper we report on the spectral analysis of the \xmm\ EPIC observations of \mkn, 
one of the brightest and
best-studied Narrow Line Seyfert 1 (NLS1) galaxies. In the X-ray region the established
characteristics of NLS1s are, a relatively steep power law spectrum and strong and rapid variability. \mkn\ is a 
particularly interesting candidate
because of its similarities with MCG-6-30-15, the archetypal relativistic broad 
Fe K line Seyfert.  
At $z=0.0129$ (with $ H_0 = 75 $~km\,s$^{-1}$\,Mpc$^{-1})$ \mkn\ has 
an X-ray luminosity (2--10~keV) which historically lies in the range
$0.5-1\times10^{43}$~erg s$^{-1}$.
The Galactic absorption column towards \mkn\ is $1.8\times10^{20}$~cm$^{-2}$
(Murphy \et\ 1996), rendering it easily visible over the whole
($\sim$0.2--12~keV) spectral band of the EPIC detectors on \xmm.

Previous X-ray observations of \mkn\ have provided a somewhat confusing picture. Though not one of the most
convincing cases, \mkn\ was included in the aforementioned
\asca\ spectral survey of Seyfert galaxies showing evidence of a relativistic broad
Fe K emission line (Nandra \et\ 1997). A separate analysis of simultaneous \rosat\ and
\asca\ observations (Leighly \et\ 1996) showed the X-ray
spectrum to be described by a power law
of index increasing strongly with flux (from $\Gamma$$\sim$1.6--2, assuming a reflection factor 
R~=~$\Omega$/2$\pi$, where $\Omega$ is the solid angle subtended by
the reflecting matter, of unity), but with only
a narrow Fe K emission line (equivalent width $\sim$100~eV at
$\sim$6.4~keV).
A later observation with \sax\ found a somewhat steeper power law
($\Gamma$$\sim$2.2), and evidence for an absorption edge at
$\sim$7.4 keV (Matt \et\ 2000). Those authors coupled the non-detection of a
narrow 6.4 keV Fe K line
with the edge detection to suggest strong reflection from moderately ionised
matter.
 
It is apparent that studies of the Fe K spectral region in NLS1s, with their characteristic steep power law spectra, were  
particularly limited by the poor sensitivity of observations above $\sim$7 keV, prior to the launch of \xmm.
In that context the existence now of two long observations of \mkn\ with \xmm\ are of particular
interest. An initial analysis of the first EPIC data set has been published by Page \et\ (2001), finding
evidence for a broad Fe K emission line. However, the main focus of that analysis was a study of spectral
variability. A first report on the second \xmm\ observation of \mkn\ is mainly concerned with the analysis and
interpretation of the RGS spectrum, but does include reference to a  
relativistically broadened
Fe K line (Mason \et\ 2003). Added importance is attached to the
latter claim by the similarity of the reported Fe K line profile to the
controversial detection of relativistic emission lines of OVIII, NVII and CVI in the soft X-ray spectrum 
of \mkn\ and MCG-6-30-15 (Branduardi-Raymont \et\ 2001; Lee \et\ 2001).

The present paper reports a new analysis of the EPIC spectral data from both observations, focussing in particular on 
the Fe K region.

\section{Observation and data reduction}

\mkn\ was observed by \xmm\ on 2000 May 20 and again on 2001 May
20-21. Useful exposure times of $\sim$60 ksec and $\sim$130 ksec were
obtained. In this paper 
we primarily use data from the EPIC 
pn cameras
(Str\"{u}der \et 2001) which have the best sensitivity of any instruments flown to date in the $\sim$6-10 keV spectral band,
relevant to a study of Fe K spectral features. However, we have confirmed that the MOS spectra are entirely consistent with the results
reported here.

The EPIC pn data were first screened with the XMM SAS v5.3 software.
X-ray events corresponding to 
patterns 0-4 (single and double pixel events) were selected.
A low energy cut of 200 eV was applied to all data
and known hot or bad pixels were removed.
We extracted source and background spectra with a circular source region of
45\arcs\ radius defined around the
centroid position of \mkn, with the background being taken from an
offset position close to the source.
The 0.2-10 keV X-ray light curve from the second observation is
reproduced as figure 1 and shows Mkn 766 in a typically active state, with factor 2 flux changes in a few ksec.
The Small Window mode was
chosen for both pn observations to minimise the effects of photon pile-up. Individual spectra 
were binned to a minimum of 20 counts per
bin, to facilitate use of the $\chi^2$ minimalisation technique in
spectral fitting. 
Response functions for spectral fitting to the \xmm\
data were generated from the SAS v5.3. 

Spectral fitting was based on the Xspec fitting package (version 11.1). All spectral fits include absorption due to the line-of-sight
Galactic column of $N_{H}=1.8\times10^{20}\rm{cm}^{-2}$, and 
fit parameters 
are given in the AGN rest-frame.
Errors are quoted at
the 90\% confidence level (e.g. $\Delta \chi^{2}=2.7$ for one
interesting parameter).

\begin{figure}
\centering
\includegraphics[width=6.3 cm, angle=270]{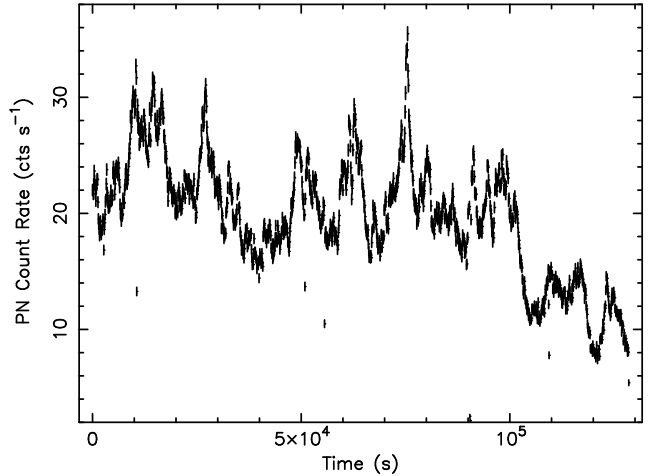}
\caption
{X-ray light curve at 0.2-10 keV from the \xmm\ pn observation of \mkn\ on 2001 May 21-22.}
\end{figure}

\begin{figure}                                                          
\centering                                                              
\includegraphics[width=5.7 cm, angle=270]{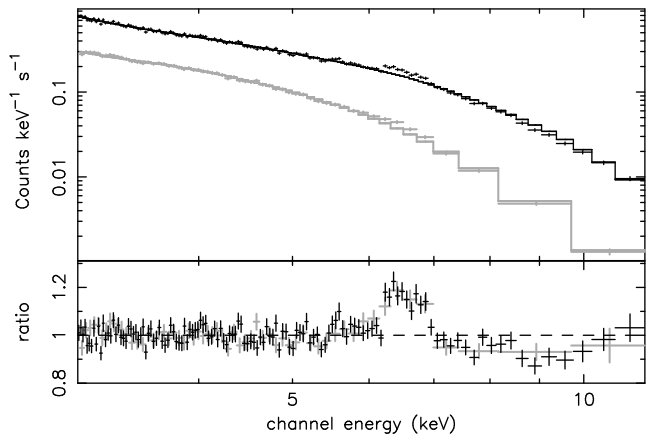}                     
\caption                                                                
{EPIC pn (black) and MOS (grey) spectral data from the 2001 May \xmm\ observation of \mkn\ compared  
with a simple power law model.   
The upper panel shows the count spectra                                 
(crosses) and the model folded through the detector                     
responses (histograms).  The lower panel shows the fit residuals.}      
\end{figure}

\section{3--11 keV spectrum}  
\subsection{Power law}
X-ray spectra of AGN are well fitted, to first order, with a power law of photon index $\Gamma$ in the range $\sim$1.6-2 for most radio
quiet AGN, with a fraction (eg NLS1) having somewhat steeper indices. The widely held view that this `hard' X-ray continuum in Seyfert
galaxies
arises by Comptonisation of thermal emission from the accretion disc in a 'hot' corona, predicts that such a power law should apply
over a broad energy
range between that of the disc photons ($\sim$25 eV) and the coronal electrons ($\sim$150 keV), a range comfortably exceeding 
the energy
band covered by the EPIC instruments on \xmm.
Determining the slope and intensity of this spectral component remains a key to the analysis of absorption and emission features
superimposed on the X-ray continuum.
To tightly constrain the value of $\Gamma$ it is valuable to have simultaneous data at higher energies, with which to determine
the amount of continuum `reflection'. In the present case there was a simultaneous \sax\ observation of \mkn\ and we have used this 
to obtain a reflection factor
of R = 1.7$\pm$0.4. Unfortunately, this can only provide an upper limit to the reflection in \mkn\ due to the presence in the 
field of view of
the \sax\ PDS instrument of the bright BL Lac object PG1218+304 (as previously noted by Matt \et\ 2000).
In our subsequent analysis we have therefore assumed a `typical' value of R = 1. 

We began our analysis of \mkn\ by fitting the \xmm\ data for the longer
2001 May observation above 3 keV, where any
effects of intervening absorption might be expected to be small, a view
supported by the analysis of the warm absorber with the high
resolution gratings on \xmm\ (Mason \et\ 2003). Since we are also interested in any gross spectral changes 
with source brightness we
excluded the final 25 ksec of this observation, where the flux level was much
lower (figure 1), and summed the spectra over the remaining $\sim$100 ksec to give our `high state' spectrum. 
Throughout this period the background rate was low and stable.

A simple power law fit over the 3--11 keV band (including cold reflection of R = 1, modelled by PEXRAV in Xspec) yielded a photon index of 
$\Gamma$$\sim$2.2,
with a broad excess in the data:model
ratio between 6--7 keV and a clear deficit above $\sim$8.7 keV (figure 2), giving an unacceptable $\chi^{2}$/dof of 1426/1190.  
 
\subsection{Fe K emission and absorption features}

To improve this fit we then added spectral components to match the most obvious features 
in the data.
The addition of a Gaussian 
emission line, with all parameters free, improved the fit
markedly (to $\chi^{2}$/dof
= 1316/1187), with a mean line energy at 6.55$\pm$0.03 keV, 1 sigma line
width 270$\pm$36 eV, and equivalent width 124$\pm$35 eV.

Given the frequent detection of a narrow `neutral' Fe K line in
other Seyfert 1 galaxies, we tested this fit for an additional narrow
Gaussian line, with energy fixed at 6.4 keV. The result was a
further improvement in the fit (to $\chi^{2}$/dof
= 1304/1186). The narrow 6.4 keV line had a (poorly constrained) equivalent width of
$\sim$40 eV, reducing
the broad, higher energy line by 
a similar
amount. The mean energy of the broad Gaussian line increased to 6.73$\pm$0.06 keV ( now consistent with the resonance
line of He-like Fe at 6.7 keV).

Visual examination of this power law plus double emission line fit showed
the deficit of flux remained at $\sim$8.5-10 keV. 
To model this feature we then added an
absorption edge with energy and optical depth as free parameters. The
result was a further significant improvement in the fit (to $\chi^{2}$/dof
= 1272/1184), with an edge energy at $\sim$8.77 keV (consistent with the He-like Fe XXV absorption edge at
$\sim$8.76 keV) and optical depth
$\sim$0.15.
Details of this parametric fit to the May 2001 pn data are 
listed in Table 1, fit 1, and the unfolded spectrum is shown in figure 3.

\begin{figure}
\centering
\includegraphics[width=6.3 cm, angle=270]{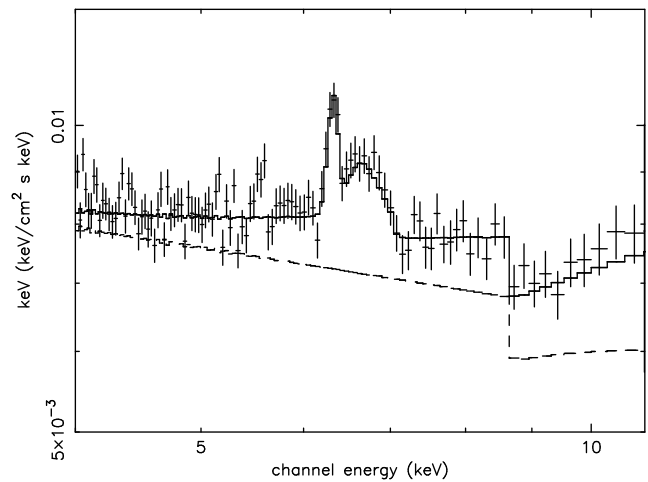}
\caption
{Unfolded spectrum illustrating the parametric model fit to the 2001 May \xmm\ observation of \mkn\ as detailed in 
Table 1, fit 1} 
\end{figure}

In summary, we find the 3-11 keV spectrum of \mkn\ to be well described by a moderately steep power law continuum, plus 
broad/ 
ionised (and probably weak/neutral) line emission, and a significant 
absorption feature above $\sim$8.7 keV. 
A physical interpretation of this fit, where the broad emission and absorption features arise in highly ionised `clouds' above
the hard X-ray continuum source, is discussed in Section 5.3. The assumed cold reflection continuum , and 
weak $\sim$6.4 keV emission line indicate separate reprocessing in near-neutral matter. 

\subsection{Reflection from an ionised disc}

Reflection from ionised matter offers an alternative, presently more conventional, origin of the broad $\sim$6.7 keV 
emission line. We therefore 
tested a model for the 2001 May \xmm\ pn data, with the hard continuum source illuminating dense 
matter (as in an
accretion disc).
We chose the XION model available in Xspec v.11.2, which includes the effects of relativistic broadening and predicts He-like Fe K line
emission for disc illumination with a relatively steep power law, as in \mkn.
The fit parameters of the model (Nayakshin \et 2001) include the source geometry, disc inclination and accretion rate through the disc.
As in the previous fit we included a narrow line at $\sim$6.4 keV as a separate component. Our trial fit used the `lamppost' geometry
with source height fixed at 5$R_{s}$, where $R_{s}$ is the Schwarzschild radius, the ratio of ionising to disc flux Fx/Fd = 0.3, solar 
abundance of Fe, and outer 
disc radius = 100$R_{s}$.
Free parameters were, the accretion rate,
incidence angle to the disc and the inner disc radius. The outcome was a reasonable fit ($\chi^{2}$/dof
= 1314/1187), with $\dot{m}$=0.5$\pm$0.2, Cos$\theta$$\sim$0.9, and $r_{in}$ = 35$\pm$6$R_{s}$. However, examination of the 
data:model
residuals again showed a strong absorption feature at $\sim$8.7 keV (figure 4). The addition of an absorption edge at $\sim$8.76 keV
($\tau$$\sim$0.14) 
improved the fit significantly (to $\chi^{2}$/dof
= 1287/1185).

Our conclusion from this `ionised reflector' fit is that the broad, $\sim$6.7 keV emission line is compatible with reflection from 
an ionised disc, with the relatively steep power law continuum of \mkn\ resulting in a `warm reflecting layer' of mainly He-like Fe,
and the innermost disc being absent (or fully ionised), in accord with the absence of a relativistic broad line. 
We note the deduced value of $\dot{m}$ in \mkn\ is consistent with 
NLS1 being thought to be accreting at a relatively
high rate (eg Pounds and Vaughan 2000).
However, we emphasise that our ionised disc fit is only illustrative, the main objective here being to see if the $\sim$8.7 keV absorption
feature is explained by (the same) reflection as produces the $\sim$6.7 keV emission line. Our fits suggest this is not the case, probably
due to the high level of scattering in the ionised skin of the disc, which predicts a broader and shallower absorption than that seen.  

\begin{table*}
\centering
\caption{Fit parameters to the 3--11 keV spectrum of \mkn\ from (1) the May 2001 observation, (2) the
May 2000 observation. 
$^a$ Rest energy of line (keV).
$^b$ Intrinsic (1 sigma) width of line (eV)
$^c$ Equivalent width of line (eV)
$^d$ Energy of absorption edge (keV)
$^e$ Optical depth of edge absorption }

\begin{tabular}{@{}lcccccccc@{}}
\hline
Fit & $\Gamma$ &6.4 keV line &\multicolumn{3}{c}{broad Fe K
line} & \multicolumn{2}{c}{absorption}
& $\chi^{2}$/dof \\

 & \ & EW$^c$ & E$^a$ & $\sigma^b$ & EW$^c$ & 
E$^d$ & $\tau^e$ &\\

\hline

1& 2.21$\pm$0.01 & 38$\pm$15 & 6.73$\pm$0.06& 160$\pm$80 &
65$\pm$20 & 8.77$\pm$0.08 & 0.15$\pm$0.03 
&1272/1184 \\

2& 2.04$\pm$0.03 & 45$\pm$35  & 6.78$\pm$0.19 & 230$\pm$150 &
70$\pm$30 & 7.2$\pm$0.15 & 0.12$\pm$0.05 &655/724 \\

\hline
\end{tabular}
\end{table*}

\begin{figure}
\centering
\includegraphics[width=6.3 cm, angle=270]{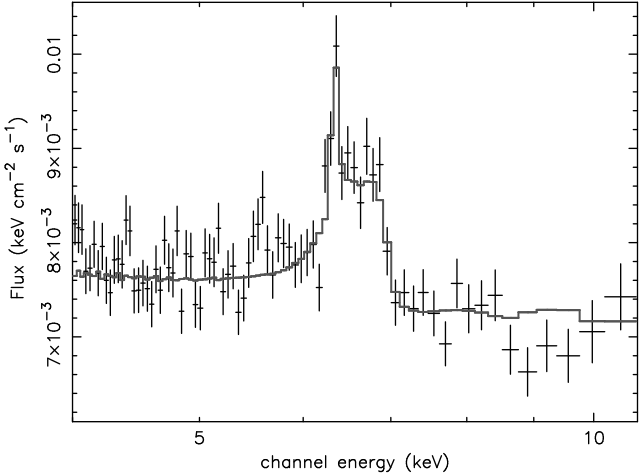}
\caption
{Unfolded spectrum illustrating the ionised disc reflection model fit to the 2001 May \xmm\ observation of \mkn\ as described in 
Section 3.3.} 
\end{figure}

\section{Comparison with the 2000 May observation}

A preliminary analysis of the pn data from the shorter 2000 May \xmm\ EPIC observation 
of \mkn\ has been
reported by Page \et\ (2001). We have re-analysed the mean EPIC spectrum over the 
first 45 ksec
of that observation when the mean 0.3-10 keV X-ray flux was only $\sim$0.5 of that a year later. We identify this as our `low state'
spectrum. Again, the most obvious spectral feature 
(figure 5) is an excess of flux at
$\sim$6-7 keV. There is also a deficit of flux at a higher energy, in this case more evident between $\sim$7--8.5 keV.
Applying the same best-fit model that we found for
the pn data from the 2001 May observation gives
the parameters in fit 2 of Table 1. The fit is very good, with a slightly flatter
power law slope (by $\sim$0.16). The Fe K line parameters are unchanged within the statistical errors,
but the absorption does appear to have moved to lower energy,  
still implying a 
substantial column
density of absorbing matter, though now in a lower ionisation state. We note this cooler absorbing matter
might demand partial covering so as not to be in conflict with the soft X-ray spectrum.
(A high reflection factor 
could provide an 
alternative
explanation, though this would be inconsistent with the weak $\sim$6.4 keV emission line.)

\begin{figure}                                                          
\centering                                                              
\includegraphics[width=5.7 cm, angle=270]{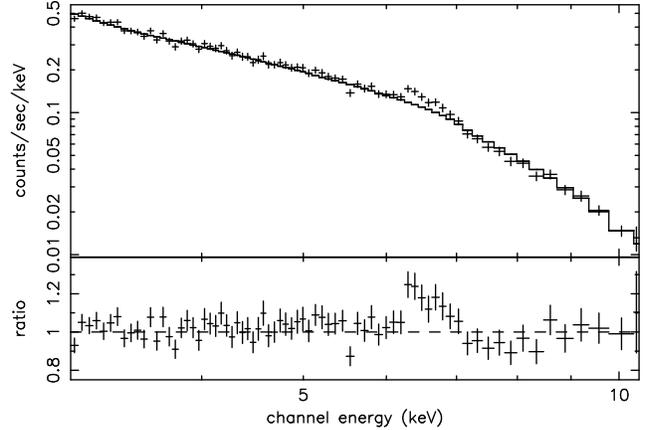}                     
\caption                                                                
{EPIC pn spectral data from the 2000 May \xmm\ observation of \mkn\ compared  
with the simple power law model described in the text.   
The upper panel shows the count spectra                                 
(crosses) and the model folded through the detector                     
responses (histograms).  The lower panel shows the fit residuals.}      
\end{figure}

\subsection{Spectral variability}

To explore the spectral variability in a model-independent way we plot in figure 6
the ratio of the high and low state spectral data. To first order this ratio rises smoothly from 10 to 1 keV,
as would be expected for a power law index increasing with flux, and then falls at lower energies (indicating a separate and less variable
component below $\sim$1 keV). However, a positive feature
between 
$\sim$7.5-8.5 keV is consistent with our parametric fits that indicate additional absorption in this band in 
the low state
spectrum. The alternative interpretation of figure 6, that the flux ratio is showing reduced variability of a broad emission line, 
seems less 
compelling from this plot.

\begin{figure}                                                          
\centering                                                              
\includegraphics[width=6.3 cm, angle=270]{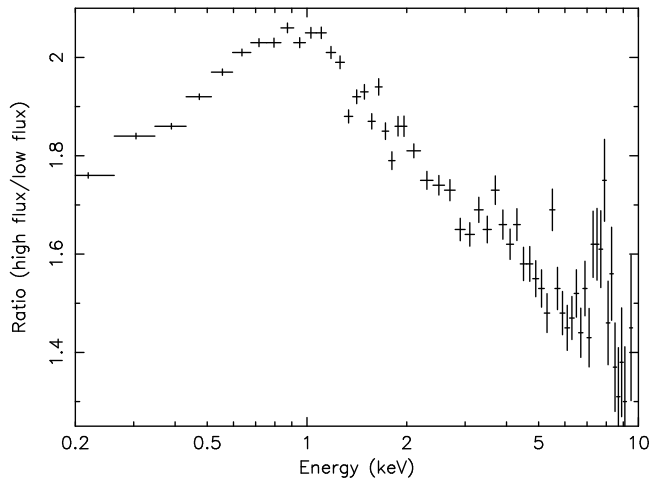}                     
\caption                                                                
{Ratio of pn spectral data from the 2001 May and
 2000 May \xmm\ observations.}      
\end{figure}

\begin{figure}                                                          
\centering                                                              
\includegraphics[width=6 cm, angle=270]{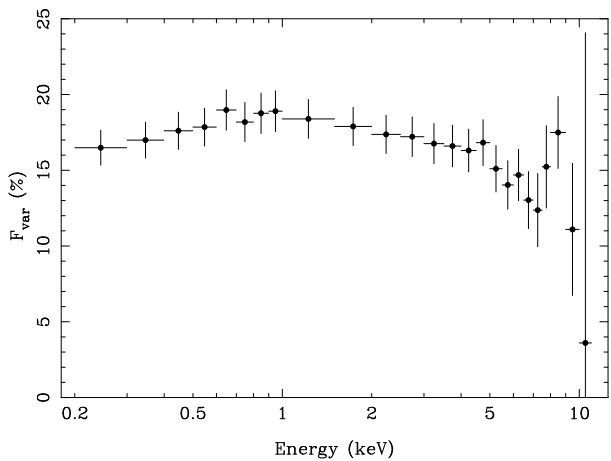}                     
\caption                                                                
{Fractional variability amplitude of the pn data from the 2001 May \xmm\ observation of \mkn}      
\end{figure}

To seek further insight on the spectral variability of 
\mkn\ we then looked for evidence of change {\it during} the longer 2001 May \xmm\ observation. The outcome is reproduced in figure 7, 
which 
shows the fractional (rms) variability amplitude, after subtraction
of the Poisson noise, as a function of 
energy (Edelson \et\ 2002, Appendix A). This plot clearly indicates a change in the degree of short-term (ie within
the 100 ksec observation) variability in the 6--9 keV band. This again might be interpreted in terms of reduced variability
in the Fe K emission line, but that would imply the line extending to $\sim$7.5 keV, which the spectral analysis does not support.
A more compelling explanation is that the enhanced short-term variability in the 7.5--9.5 keV energy band is due to
variable, ionised absorption. The implication is that ionised matter is moving in front of the continuum source, or is changing in
ionisation state, within the observing time span.

\subsection{Flare and non-flare spectrum}

Given the above evidence for rapid spectral changes in the Fe K band we then examined the individual spectra in six 20 ksec intervals 
across
the complete May 2001 observation. Spectral variability was indicated but without any clear pattern being evident. We then made a
different cut, separating the data with an overall count rate (see figure 1) above and below 21 counts/sec, thereby seeking to resolve
changes associated with the strong flux peaks. The outcome of this 
split is
illustrated in figure 8, which shows the data:model ratio plots for the `flare' and `non-flare' spectra, against
the 
best fit power law. Two differences stand out. First, the broad Fe K emission is significantly {\it weaker} in the `flare' data; second, 
the ionised
absorption
appears at a significantly higher energy. An alternative view of these spectral changes is shown in  
figure 9, plotting the ratio of flare to non-flare data. The implication of this plot is
that the ionised absorption edge does indeed move to a higher energy during the flares, while increased absorption may also be responsible
for depressing the broad emission line. The inference from this interpretation is that highly ionised matter is injected into the line of
sight during the flares. On this view the shift in the absorption edge energy could reflect an increased level of ionisation (He-like to
H-like Fe) or be a measure of the injection velocity of (He-like) matter. If the marked decrease in the strength of the $\sim$6.7 keV
emission is due to resonant line absorption, the implied opacity requires a large velocity dispersion and/or a range of ionisation states.
Higher resolution spectra will be required to address this question.

\begin{figure}                                                          
\centering                                                              
\includegraphics[width=5.7 cm, angle=270]{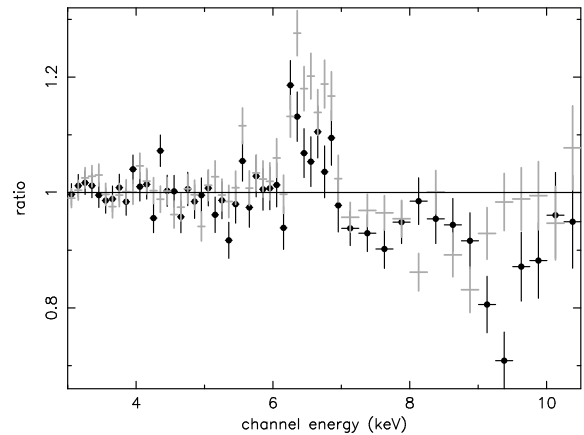}                     
\caption                                                                
{Spectral data from the 2001 May \xmm\ observation of \mkn\ split into `non-flare' (grey) and `flare' (black) components, as 
described in the text, and plotted as a ratio of data to a simple power law.}      
\end{figure}

\begin{figure}                                                          
\centering                                                              
\includegraphics[width=5.7 cm, angle=270]{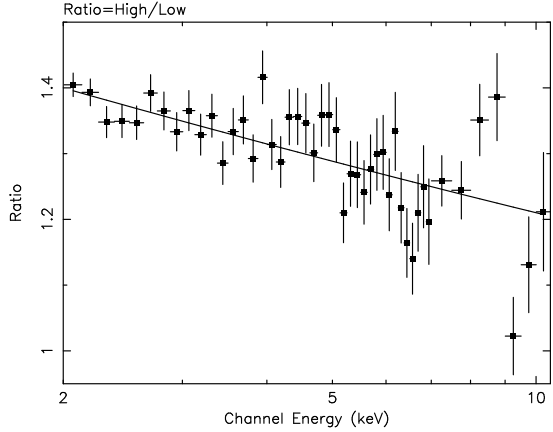}                     
\caption                                                                
{Plot of the ratio of `flare' to `non-flare' spectral data from the 2001 May observation.}      
\end{figure} 

To quantify the visible differences in the `flare' and `non-flare' spectra we then fitted each data set with the parametric models
used for the full 2001 May observation (Table 1, fit 1). The fit details are given in Table 2. In summary, the non-flare spectrum is
similar to the mean for the complete data, but the flare spectrum has the most obvious absorption edge at a higher energy of 
$\sim$9.15 keV (close to the H-like Fe edge at 9.28 keV). The emission line fit
includes an unchanged 6.4 keV line, but substantially weaker emission at $\sim$6.7 keV. The photon flux value for the ionised emission 
line is
1.3$\pm$0.3$\times 10^{-5}$~cm$^{-2}$~s$^{-1}$ in the non-flare spectral fit, falling to 5.5$\pm$2.7 $\times 10^{-6}$~cm$^{-2}$~s$^{-1}$
in the flare spectrum.
It seems likely that this reduction in the observed ionised line emission is 
due to absorption not fully resolved
by the pn detector. 

\begin{figure}
\centering
\includegraphics[width=6.3 cm, angle=270]{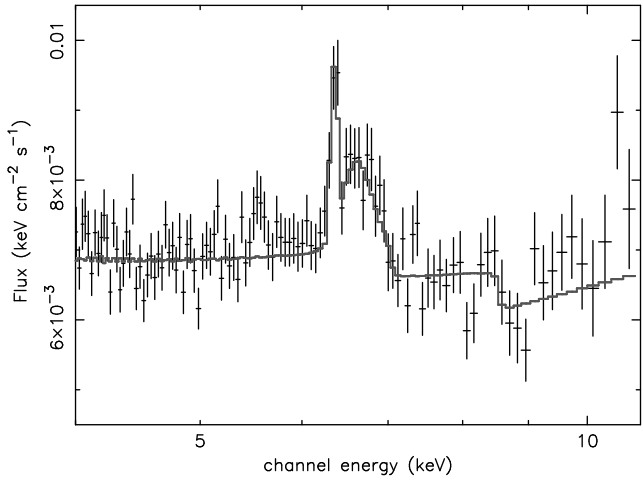}
\caption
{Unfolded spectrum illustrating the parametric model fit to the 2001 May non-flare spectrum of \mkn\ as detailed in Table 2, fit 1.} 
\end{figure} 

\begin{figure}
\centering
\includegraphics[width=6.3 cm, angle=270]{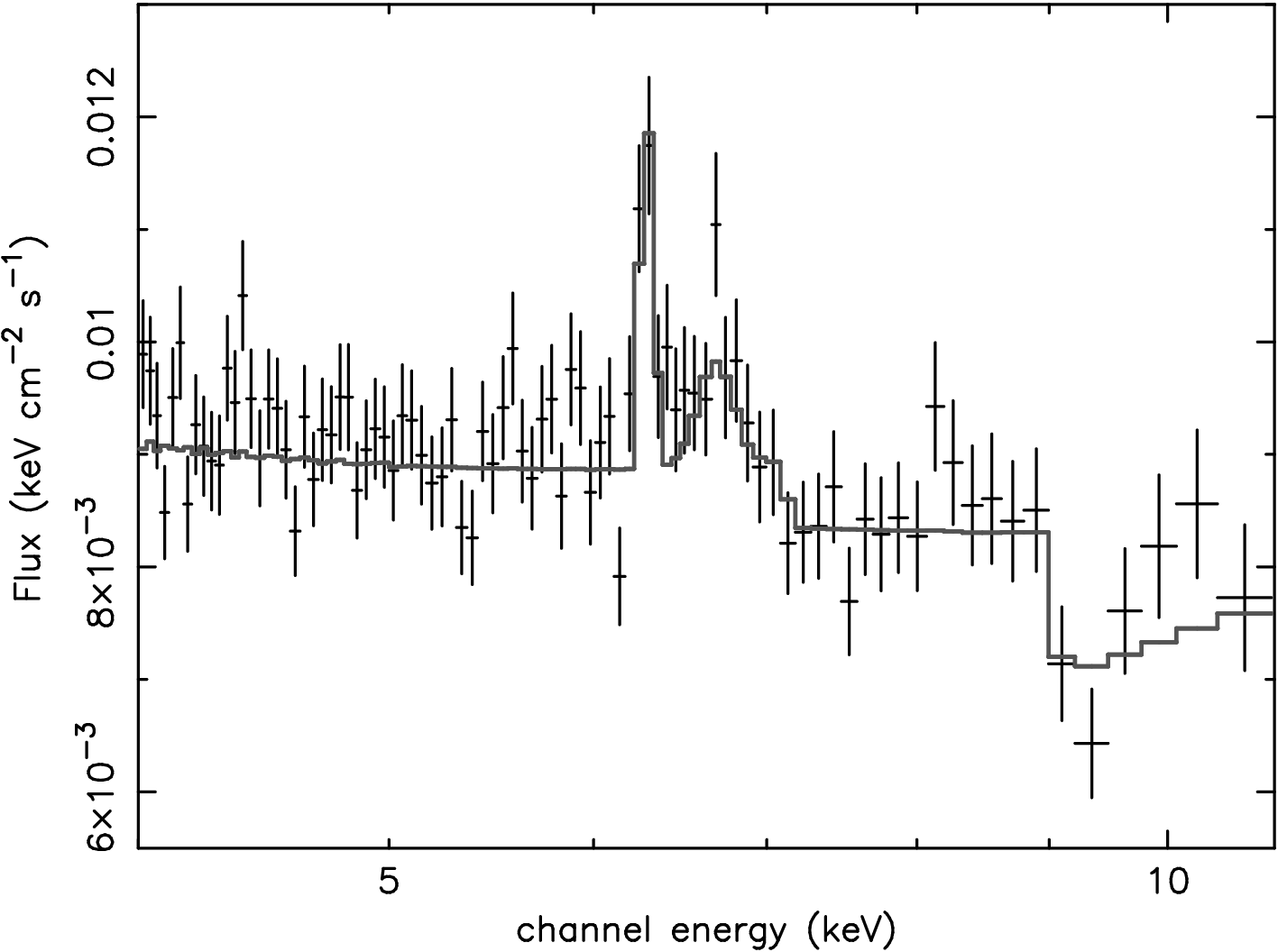}
\caption
{Unfolded spectrum illustrating the parametric model fit to the 2001 May flare spectrum of \mkn\ as detailed in Table 2, fit 2.} 
\end{figure}

\begin{table*}
\centering
\caption{Fit parameters to (1) the `non-flare' and (2) the `flare' spectral data at 3-11 keV for the 2001
May observation of \mkn.
$^a$ Rest energy of line (keV).
$^b$ Intrinsic (1 sigma) width of line (eV)
$^c$ Equivalent width of line (eV)
$^d$ Energy of absorption edge (keV)
$^e$ Optical depth of edge absorption }

\begin{tabular}{@{}lcccccccc@{}}
\hline
Fit & $\Gamma$ &6.4 keV line &\multicolumn{3}{c}{broad Fe K
line} & \multicolumn{2}{c}{absorption}
& $\chi^{2}$/dof \\

 & \ & EW$^c$ & E$^a$ & $\sigma^b$ & EW$^c$ & 
E$^d$ & $\tau^e$ &\\

\hline

1& 2.16$\pm$0.02 & 40$\pm$15 & 6.67$\pm$0.08& 190$\pm$90 &
87$\pm$30 & 8.66$\pm$0.12 & 0.12$\pm$0.03 
&1180/1098 \\

2& 2.25$\pm$0.06 & 32$\pm$18  & 6.8$\pm$0.15 & 75$\pm$25 &
25$\pm$20 & 9.15$\pm$0.09 & 0.23$\pm$0.06 & 731/755 \\

\hline
\end{tabular}
\end{table*}

\section{Discussion}

Our analysis of the \xmm\ EPIC data on \mkn\ emphasises again the 
often subtle effects of
absorption and reflection in fitting broad-band X-ray spectra.
Although current observing facilities still have relatively poor sensitivity
above $\sim$7 keV, this spectral region is a unique diagnostic of a
potentially large column density of ionised matter in which all abundant elements
lighter than Fe may be fully stripped. Neglect of such
absorption can lead to an overestimate of broad Fe K line emission.
More generally, the present analysis of the \xmm\ spectrum of \mkn\ strengthens
an emerging view that the extreme
red wing, indicated to be a common feature of Seyfert 1 galaxies in
\asca\ data (eg Nandra \et\ 1997), is probably less strong and less common than previously reported.   
  
\subsection{Fe K line emission}
  
The \xmm\ observations of \mkn\ do, however, indicate complex spectral structure, with the 6--7 keV 
energy band containing emission from both neutral and ionised Fe.
Our data cannot uniquely resolve the
emission components in \mkn, but a plausible 
solution includes lines with energies compatible with near-neutral and He-like Fe.
What is notable is the relative weakness of both lines. In the case of
the 6.4 keV line, the equivalent width of $\sim$40 eV is barely
compatible with
the previously reported cold reflection, R $\sim$1. 
An explanation of the weak line emission in terms of an 
unusually low abundance of Fe 
appears at odds with optical spectra of \mkn\ and NLS1 in general. 
An interesting alternative, for the 6.4 keV line, could be that the relevant fluorescing matter is moderately 
ionised ($\xi$ in the range 100-500), with resonance absorption and subsequent Auger destruction of the line 
photons(Matt \et\ 1993) - and would imply that
a significant fraction of the corresponding reflection occurs closer to the hard X-ray source than 
the
molecular torus.

That much of the reprocessing matter is highly ionised is apparent from the observed He-like
emission and absorption features. Paradoxically the weak He-like emission line is at odds with the higher
fluorescent yield for that ion. The implication, for a reflection model, is that emergence through a 
substantial
optical depth of fully ionised matter is removing
flux from the detected line cores. 
This ionised matter could exist in the form of a `skin' to the inner accretion disc, an effect included in the ionised disc
model we used. However, the visual evidence for substantial absorption by ionised Fe in \mkn\ suggests that 
line emission from the accretion disc may also be attenuated in passage through He-like Fe lying above the hard
X-ray source. A more radical alternative is that both Fe K absorption and (re-)emission arises mainly in such `external' matter.

\subsection{Line-of-sight absorption}

A broad absorption trough is clearly visible in the 2001 May pn data (figure 2). Fitting this feature with an absorption edge finds 
a `rest' energy of 8.8$\pm$0.1 keV, close to that of
He-like Fe (8.76 keV). With a threshold absorption cross-section of $2\times$$10^{-20}$~cm$^{2}$ (Verner and Yakovlev, 1995) the 
measured optical depth of 0.12$\pm$0.03 implies 
a column density of  
order $6\times$$10^{18}$~cm$^{-2}$ of He-like Fe (or a hydrogen equivalent column for solar abundance of order
$2\times$$10^{23}$~cm$^{-2}$). 
We note the recent detection of an apparently similar ionised absorption feature in the X-ray spectrum of a BAL QSO 
(APM 08279+5255),
alternatively identified with the absorption edge of Fe XV-XVIII (Hasinger \et\ 2002), or strongly blue-shifted resonance absorption
lines of Fe XXV or XXVI (Chartas \et\ 2002), also implies a substantial line-of-sight column of highly ionised matter. 
Earlier, a long \asca\ observation of NGC 3516 yielded evidence for absorption superimposed on the `red wing' of a broad Fe K
emission line (Nandra \et\ 1999).
Our analysis of the \mkn\ spectrum from both \xmm\ observations suggests further absorption (additional to the main edge)
in the 7--10 keV band,
and we recall the \sax\ observation of \mkn\ also found evidence for
an absorption edge at $\sim$7.4 keV with an optical depth ($\tau$$\sim$0.26), greater than would be expected from a simple 
disc reflection model (Matt \et\ 2000).
Other reports of surprisingly strong absorption in the Fe K band, mainly from \xmm\ observations,
include the NLS1 1H 0707-495 (Boller \et\ 2002), and the luminous low redshift QSO PDS 456 (Reeves \et\ in preparation).

We conclude that the Fe K spectral band in many AGN may be affected by substantially more absorption 
than has generally been recognised. More complex modelling may be necessary, and we note, for example, 
that inner-shell transitions in 
a wide range of ionisation states, analogous to those
detected in the soft X-ray band (eg Sako \et\ 2001), have recently been discussed in relation to the structure of 
the Fe K edge in
ionised matter by Palmeri \et\ (2002).  

\subsection{Absorption and re-emission in ionised clouds}

Although the first suggestion of X-ray `reflection' in AGN was made in terms of reprocessing in dense,
accreting clouds (Guilbert and Rees 1988), since the predicted spectral features became well established
(Pounds \et\ 1990, Nandra and Pounds 1994), the usual interpretation has assumed the scattering and fluorescence to arise from the
putative accretion disc.
On this view our hard X-ray spectrum of \mkn\ would invoke 2-component reflection, 
from distant/cold and nearby/ionised matter (perhaps the molecular torus and inner accretion disc, respectively). Compton
scattering in a fully ionised `skin' overlying the inner disc could then explain the relative
weakness and width of the ionised 6.7 keV emission line, particular if allowance was made for line blending. [The
absence of a `red wing' would suggest the innermost disc was fully ionised, or absent.] 
However, as noted in Section 3.3, the relatively `sharp' ionised absorption edge appears to be inconsistent with this simple 
reflection
model, as is the weakness of the $\sim$6.4 keV line (unless
the abundance of Fe is much less than `solar').

The analysis presented in this paper
has demonstrated that taking account of such absorption can weaken the evidence for reprocessing to be taking place in a
region of strong gravity (the innermost accretion disc). At the same time indications of a large
column density of ionised gas in the line of sight to the continuum X-ray source offers another candidate for the
reprocessing matter. 
We briefly explore such an alternative scenario to explain the X-ray spectral features in 
\mkn, in which the hard X-ray continuum flux emerges through a distribution of ionised matter (`clouds') at a small radial distance. 
Such clouds will see a much higher ionising flux and have higher
velocities, for example, than the optical broad line clouds.

The measured optical depth of the $\sim$8.7 keV edge implies a cloud column density of He-like Fe of 
$\sim 6 \times 10^{18}$~cm$^{-2}$. The photon flux absorbed 
by the He-like edge is $\sim 8 \times 10^{-5}$~cm$^{-2}$~s$^{-1}$.
Assuming a fluorescent yield of 0.75, re-emission from a spherical distribution of such clouds would match the observed
6.7 keV emission line flux of
$\sim 10^{-5}$~cm$^{-2}$~s$^{-1}$ for a
covering factor of $\sim$0.2. Apparently such re-emission could 
make a significant contribution to the observed He-like Fe K emission. 

It is interesting to extend this simple assessment of a circumnuclear cloud model by interpreting 
the observed line width in terms of the velocity dispersion of the emitting material (ie in the traditional way for the broad 
optical lines),
while recalling that the small equivalent width of the $\sim$6.7 keV line does suggest
significant Comptonisation.
However, on the assumption that the `core' line broadening is primarily Doppler, then the measured
line width of
$\sigma$ = 200 eV, or $\sim$22000 km s$^{-1}$ at FWHM, equates to a radial distance of $\sim10^{14}$~cm 
for a black hole mass
for \mkn\ of $\sim 3\times 10^{6}$ $\Msun$.
Assuming a cloud radius of $\sim 3 \times 10^{13}$~cm, the observed column density then gives a mean cloud density
of $\sim 7 \times 10^{9}$~cm$^{-3}$.
A rough consistency check on these parameters can be made by calculating the ionisation parameter for
such a cloud irradiated by the ionising continuum source in \mkn.
Integrating the measured luminosity above the Fe K edge we find log $\xi$ = $L/nr^2$ $\sim$3.8, confirming that if such a cloud 
existed then 
it would indeed be
ionised to a level where He-like Fe dominates.

The purpose of the above estimates was to point out that ionised matter lying above the continuum X-ray source may contribute significantly
to the 
observed Fe K emission features. However, what appears more certain is that such matter will impose significant absorption on the emerging
spectrum. Our evidence suggests this absorption varies on both the long term (ie between the 2000 and 2001 \xmm\ observations), and short
term. The degree of short-term variability (ie within
the 100 ksec observation) in the 6--9 keV band is of particular interest.
Interpreting the data shown in figure 7 in terms of variable ionised absorption, the implication is that ionised flare ejecta are 
moving in
front of the hard X-ray source, or are
changing in
ionisation state, from He-like to H-like Fe, within a timescale of a few ks. The former interpretation provides a further constraint on our cloud
model.
For a cloud of radius $3\times10^{13}$~cm to occult a point source of hard X-radiation in, say 
$\sim 2\times 10^{4}$ sec, requires a transverse velocity of $\sim$15000 km s$^{-1}$, consistent with the velocity deduced
from the X-ray line width.
A possible origin for ionised matter moving at such high velocities is suggested by the association of the observed spectral changes with 
X-ray flaring. If the energy shift of the primary absorption edge is interpreted in terms of an outflow
of flare ejecta, the corresponding velocity is $\sim$15000 km s$^{-1}$, at least an intriguing coincidence. 

We note, finally, that if the above conjecture is extended to include Fe K shell absorption by line-of-sight matter over a wider range 
of ionisation
states, as the \xmm\ spectra suggest, then this concept of partial covering may need to be invoked to explain
the continuing visibility of soft X-ray emission from \mkn.

\subsection{Are powerful X-ray flares and flare ejecta likely in \mkn\ ?}

The background-subtracted light curve of the 2001 May \xmm\ observation is shown in figure 1. It is clearly
highly variable, a characteristic of NLS1, but in a non-random manner. Indeed, similar variability in the
first \xmm\ observation of \mkn\ led Boller \et\ to suggest a periodicity. We have chosen instead to characterise
the variability as consisting of a sequence of `flares` and draw attention to the similarity with the
X-ray light curve of the luminous QSO PDS 456. In that case Reeves \et\ 2002 have suggested a model in which the
the hard X-ray pulses are coherent events arising close to the SMBH, and involving a sequence (avalanche) of
magnetic flares in which energy is deposited into the coronal electrons by magnetic reconnection. 
A simple scaling of the arguments presented in Reeves \et\ 2002, for a black hole of $M \sim 3 \times 10^{6}\Msun$ and 
$\dot{m}$=0.3, perhaps appropriate to \mkn, yields a kinetic energy flux through the inner accretion disc of $\sim10^{44}$~erg  
s$^{-1}$. For a substantial fraction of this energy to rise into the corona in the form of buoyant magnetic flux tubes
requires magnetic fields
of order $B \sim 1.5 \times 10^{4}$~G to be generated within the inner accretion disc. Hence the total energy built up in the
corona could 
be $> B^2 (3R_s)^3/8 \pi \sim 10^{45}$ erg.
This energy reservoir is comparable to the energy in a large scale fluctuation in the \mkn\ X-ray light curve.
Again, following the arguments in Reeves \et\ 2002, the Alfven time in the corona of \mkn\ would be
$\sim$300 s, allowing individual flares to occur in $\sim 10^3$ s.
With the energy release in an single flare (Di Matteo 1998) $\sim      
10^{39}$~erg s$^{-1}$, a cascade or avalanche of a few 1000 individual flares could then explain the observed fluctuations in \mkn. 
As pointed out in Reeves \et 2002, such large amplitude, coherent x-ray fluctuations may be a further signature of AGN accreting 
at a relatively high rate.

\subsection{The X-ray spectrum of \mkn, a high accretion rate AGN}

The average observed flux in the 2--10~keV band during the 2001 May
20-21 observation 
was $2 \times       
10^{-11}$~erg s$^{-1}$  cm$^{-2}$, corresponding to a    
luminosity of $\sim 7 \times 10^{42}$~erg s$^{-1}$. The bolometric   
luminosity is then likely to be of order $\sim10^{44}$~erg  
s$^{-1}$. Assuming a black 
hole mass for \mkn\ in the range M $\sim10^{6}-10^{7}\Msun$ this luminosity implies
an accretion rate of $\sim$ 10--100 percent of the Eddington rate,
in contrast to the value of only
4 percent reported for the archetypal BLS1 NGC 5548 (Pounds \et\ 2003).
We have argued previously that a high accretion rate may be the key to the characteristic X-ray properties of 
Narrow Line
Seyfert 1 galaxies (eg Pounds and Vaughan 2000), to which now might be added Fe K absorption (and possibly significant emission) due to
highly ionised circumnuclear matter.   
  
\section{Conclusions}

(1) When due account is taken of absorption in the 7--10 keV band we find there is no requirement for the
previously reported strong, relativistically broad Fe K emission line in the \xmm\ spectrum of \mkn. The residual
line emission is relatively weak, with probable components from both He-like and near-neutral Fe. 

(2) Evidence of a substantial column density of highly ionised matter in the line-of
sight to the hard X-ray source, adds a further component to the growing complexity of matter in the 
vicinity of the (once considered bare)
Seyfert 1 nucleus. 

(3) We sketch a
 model in which X-ray absorption and re-emission arise in 
circumnuclear matter, existing in the form of ionised clouds (or ejecta) close to the SMBH.

(4) Short-term variability in both Fe K absorption and emission features is found
 to be linked with large scale
X-ray flaring. 

\section*{ Acknowledgements }
The results reported here are based on observations obtained with \xmm, an ESA science mission with
instruments and contributions directly funded by ESA Member States and
the USA (NASA).
The authors wish to thank the SOC and SSC teams for organising the \xmm\
observations and initial data reduction and Simon Vaughan for several stimulating discussions.

\end{document}